\documentstyle[aps,prd,eqsecnum,epsf,twocolumn]{revtex}
\newcommand{\be}{\begin{equation}}
\newcommand{\ee}{\end{equation}}
\newcommand{\bea}{\begin{eqnarray}}
\newcommand{\eea}{\end{eqnarray}}
\newskip\humongous \humongous=0pt plus 1000pt minus 1000pt
\def\caja{\mathsurround=0pt}
\def\eqalign#1{\,\vcenter{\openup1\jot \caja
    \ialign{\strut \hfil$\displaystyle{##}$&$
    \displaystyle{{}##}$\hfil\crcr#1\crcr}}\,}
\newif\ifdtup

\relax 
\title{Some Dynamical Effects of the Cosmological Constant}
\author{M. Axenides, E. G. Floratos and L. Perivolaropoulos}
\address{Institute of Nuclear Physics, \\
National Centre for Scientific Research ``Demokritos N.C.S.R.'',\\
Athens, Greece\\ e--mail: {\tt axenides@mail.demokritos.gr, \ 
floratos@mail.demokritos.gr, \ leandros@mail.demokritos.gr}} 

\date{\today}
\draft

\begin{document}

\maketitle

\begin{abstract}
Newton's law gets modified in the presence of a cosmological 
constant by a small repulsive term (antigarvity) that is 
proportional to the distance. Assuming a value of the cosmological 
constant consistent with the recent SnIa data ($\Lambda \simeq 
10^{-52} m^{-2}$) we investigate the significance of this term on 
various astrophysical scales. We find that on galactic scales or 
smaller (less than a few tens of kpc) the dynamical effects of the 
vacuum energy are negligible by several orders of magnitude. On 
scales of 1Mpc or larger however we find that vacuum energy can 
significantly affect the dynamics. For example we show that the 
velocity data in the Local Group of galaxies correspond to 
galactic masses increased by 35\% in the presence of vacuum 
energy. The effect is even more important on larger low density 
systems like clusters of galaxies or superclusters. 

\end{abstract}

\pacs{PACS:  }

\section{Introduction}
Almost two years ago two groups (the Supernova Cosmology Project 
\cite{scpc99} and the High-Z Supernova Team 
\cite{hz98,Riess:1998cb} presented evidence that the expansion of 
the universe is accelerating rather than slowing down. These 
supernova teams have measured the distances to cosmological 
supernovae by using the fact that the intrinsic luminosity of Type 
Ia supernovae, while not always the same, is closely correlated 
with their decline rate from maximum brightness, which can be 
independently measured.  These measurements combined with redshift 
data for these supernovae has led to the prediction of an 
accelerating universe. A non-zero and positive cosmological 
constant $\Lambda$ with \be \Lambda \simeq 10^{-52} m^{-2}  
\label{valcc} \ee  could produce the required repulsive force to 
explain the accelerating universe phenomenon. A diverse set of 
other cosmological observations also compellingly suggest that the 
universe posesses a nonzero cosmological constant corresponding to 
vacuum energy density of the same order as the matter energy 
density\cite{Weinberg:1996xe,Krauss:1995yb,Carroll:1992mt}.  

In addition to causing an acceleration to the expansion of the 
universe the existence of a non-zero cosmological constant would 
have interesting gravitational effects on various astrophysical 
scales\cite{Ostriker:1995su}. For example it would affect 
gravitational lensing statistics of extragalactic 
surveys\cite{Quast:1999fh}, large scale velocity 
flows\cite{Zehavi:1999fm} and there have been some claims that 
even smaller systems (galactic\cite{Whitehouse:1999rs} and 
planetary\cite{ct98}) could be affected in an observable way by 
the presence of a cosmological constant consistent with 
cosmological expectations. Even though some of these claims were 
falsified\cite{Wright:1998bc,Neupane:1999hr,Roberts:1987ch} the 
scale dependence of the dynamical effects of vacuum energy remains 
an interesting open issue. 
 
The effects of the cosmological constant on cosmological scales 
and on local dynamics can be obtained from the Einstein equations 
which in the presence of a non-zero cosmological constant are 
written as 
\be
R_{\mu \nu}-{1\over 2} g_{\mu \nu} R = 8 \pi G T_{\mu \nu} - 
\Lambda g_{\mu \nu} 
 \label{einst}\ee These equations imply the Friedman 
equation \be {\dot R}^2={{8 \pi G}\over 3} R^2 \rho_M - k c^2 + 
{{\Lambda R^2}\over 3}\equiv {{8 \pi G}\over 3} R^2 (\rho_M + 
\rho_\Lambda) - k c^2
 \label{fried} \ee  where the vacuum energy density 
$\rho_\Lambda$ is defined as $\rho_\Lambda \equiv \Lambda /8\pi 
G$. Since the vacuum energy does not scale with redshift it is 
easily seen from eq. (\ref{fried}) that it can cause acceleration 
(${\ddot R}>0$) of the universe expansion. 

The observational evidence for accelerating expansion along with 
constraints on the matter density as derived from dynamical 
measurements of galaxies and clusters, and additional constraints 
from the anisotropies of the cosmic microwave 
background\cite{White:1998fn}, lead to consistent picture with 
${{\rho_\Lambda} \over {\rho_M}}\simeq 2$, with the total energy 
density  approximately equal to the critical density necessary to 
solve (\ref{fried}) with k=0 ($\rho_\Lambda + \rho_M\simeq 
\rho_c$). 

However, the vacuum energy implied from eq. (\ref{valcc}) 
($10^{-10} erg/cm^3$) is less by many orders of magnitude than any 
sensible estimate based on particle physics. In addition, the 
matter density $\rho_M$  and and the vacuum energy $\rho_\Lambda$  
evolve at different rates, with $\rho_M / \rho_\Lambda \simeq 
R^{-3}$ and it would seem quite unlikely that they would differ 
today by a factor of order unity. Interesting attempts have been 
made during the past few years to justify this apparent fine 
tuning by incorporating evolving scalar fields 
(quintessence\cite{Zlatev:1999tr}) or probabilistic arguments 
based on the anthropic principle\cite{Garriga:1999bf}). 

In addition to the prediction for accelerating universe, the 
presence of a non-zero cosmological constant  also affects the 
form of gravitational interactions. The generalized spherically 
symmetric vacuum solution of eq. (\ref{einst}) may be written as 
\be ds^2=A(r)c^2 dt^2 -dr^2 /A(r) - r^2(d\theta^2 + \sin^2\theta 
d\phi^2) \label{sds}\ee where $A(r)=1-2GM/c^2 r - \Lambda r^2/3$. 
This metric is known as the Schwarzschild-de-Sitter 
metric\cite{Gibbons:1977mu} and describes a space that is not 
asymptotically flat but has an asymptotic curvature induced by the  
vacuum energy corresponding to $\Lambda$. In the weak field limit 
we may use the Schwarzschild-de-Sitter metric to find the 
corresponding Newtonian potential $\phi$ \be g_{00}=A(r)=1+2 
\phi/c^2  \label{g00}\ee which leads to \be \phi =  {{G M}\over r} 
+ {{\Lambda c^2 r^2}\over 6} 
 \label{ptl}\ee This generalized Newtonian 
potential leads to a gravitational interaction acceleration  \be 
g=-{{GM}\over r^2} + {{\Lambda c^2}\over 3} r 
 \label{acc}\ee This generalized 
force includes a repulsive term \be g_r = {{\Lambda c^2}\over 3} r 
 \label{rep}\ee which is expected to dominate at distances larger than \be r_c 
= ({{3 G M}\over {\Lambda c^2}})^{1\over 3}\simeq 10^2 ({{{\bar 
M}_1}\over {{\bar \Lambda_{52}}}})^{1\over 3} pc \simeq 2\times 
10^7 ({{\bar M}_1\over {{\bar \Lambda_{52}}}})^{1\over 3} AU
 \label{crdist}\ee 
where ${\bar M}_1$ is the mass within a sphere of radius $r_c$  in 
units of solar masses $M_\odot = 2 \times 10^{30} kg$ and  ${\bar 
\Lambda_{52}}$ is the cosmological constant in units of $10^{-52} 
m^{-2}$.  

The question we address in this paper is the following: `What are 
the effects of the additional repulsive force $g_r$ on the various 
astrophysical scales?' This issue has been addressed in the 
literature for particular scales. For example it was 
shown\cite{Wright:1998bc} that the effects of this term in the 
solar system could only become measurable (by modifying the 
perihelia precession) if the cosmological constant were fourteen 
orders of magnitude larger than the value implied by the SnIa 
observations. 

A recent study\cite{Whitehouse:1999rs} has also addressed this 
issue for galactic scales. In that study an attempt was made to 
explain the flat rotation curves of galaxies without the existence 
of dark matter. It was found that a cosmological constant $\Lambda 
\simeq 10^{-48}m^{-2}$ can account for the flat rotation curves of 
M33 and other galaxies  and is consistent with a theoretical value 
obtained from the Extended Large Number Hypothesis. This value of 
$\Lambda$ however is four orders of magnitude larger than that of 
eq. (\ref{valcc}) indicated by supernova and other cosmological 
observations. 

In the next section it will be shown that the vacuum energy 
required to close the universe (eq. (\ref{valcc})) has negligible 
effects on the dynamics of galactic scales (few tens of kpc). The 
dynamically derived mass to light ratios of galaxies obtained from 
velocity measurements on galactic scales are modified by less than 
$0.1\%$ due to the vacuum energy term of eq. (\ref{valcc}). This 
is not true however on cluster scales or larger. Even on the 
scales of the Local Group of galaxies (about 1Mpc) the 
gravitational effects of the vacuum energy are significant. We 
show that the dynamically obtained masses of M31 and the Milky Way 
must be increased by about 35\% to compensate the repulsion of the 
vacuum energy of eq. (\ref{rep}) and produce the observed relative 
velocity of the members of the Local Group. The effects of vacuum 
energy are even more important on larger scales (rich cluster and 
supercluster). 

\section{Scale Dependence of Antigravity} 

In order to obtain a feeling of the relative importance of 
antigravity vs gravity on the various astrophysical scales it is 
convenient to consider the ratio of the corresponding two terms in 
eq. (\ref{acc}). This ratio $q$ may be written as \be q={{\Lambda 
c^2 r^3}\over {3 G M}} \simeq 0.5 \times 10^{-5} {{{\bar 
\Lambda}_{52}{\bar r}_1^3}\over {{\bar M}_1}}  \label{qrat}\ee  
where ${\bar r}_1$ is the distance measured in units of $pc$. For 
the solar system (${\bar r}_1\simeq 10^{-5}$, ${\bar M}_1 = 1$) we 
have $q_{ss}\simeq 10^{-20}$ which justifies the fact that 
interplanetary measures can not give any useful bound on the 
cosmological constant. 

For a galactic system (${\bar r}_1\simeq 10^{4}$, ${\bar M}_1 = 
10^{10}$) we have $q_g \simeq 5 \times 10^{-4}$ which indicates 
that up to galactic scales the dynamical effects of the 
antigravity induced by $\Lambda$ are negligible. On a cluster 
however (${\bar r}_1\simeq 10^{7}$, ${\bar M}_1 = 10^{14}$) we 
obtain $q_c \simeq O(1)$ and the gravitational effects of the 
vacuum energy become significant. This will be demonstrated in a 
more quantitative way in what follows.  

The precessions of the perihelia of the planets provide one of the 
most sensitive Solar System tests for the cosmological constant. 
The additional precession due to the cosmological constant can be 
shown\cite{Wright:1998bc} to be \be \Delta \phi_\Lambda = 6\pi q 
\; rad/orbit  \label{addprec}\ee where $q$ is given by eq. 
(\ref{qrat}).  For Mercury we have ${\bar r}_1 \simeq 10^{-6}$ 
which leads to $q_{mc} \simeq 10^{-23}$ and $\Delta \phi_\Lambda 
\simeq 10^{-22} rad/orbit $. The uncertainty in the observed 
precession of the perihelion of Mercury is $0.1''$ per century or 
$\Delta \phi_{unc} \simeq  10^{-9} rad/orbit $ which is $13$ 
orders of magnitude larger than the one required for the detection 
of a cosmologicaly  interesting value for the cosmological 
constant. The precession per century\footnote{The angular velocity 
is smaller for distant planets and therefore the precession per 
century does not scale like the ${\bar r}_1^3$ as the precession 
per orbit does} scales like ${\bar r}_1^{3/2}$ and therefore the 
predicted additional precession per century for distant planets 
(${\bar r}_1 (Pluto) \simeq 10^2 {\bar r}_1 (Mercury)$)  due to 
the cosmological constant increases by up to 3 orders of 
magnitude.  It remains however approximatelly 10 orders of 
magnitude smaller than the precession required to give a 
cosmologically interesting detection of the cosmological constant 
even with the best quality of presently available observations.  
It is therefore clear that since the relative importance of the 
gravitational contribution is inversely proportional to the mean 
matter density on the scale considered, a cosmological constant 
could only have detectable gravitational effects on scales much 
larger than the scale of the solar system. 

On galactic scales, the rotation velocities of spiral galaxies as 
measured in the 21cm line of neutral hydrogen comprise a good set 
of data for identifying the role of the vacuum energy on galactic 
scales. This is because these velocity fields usually extend well 
beyond the optical image of the galaxy on scales where the effects 
of $\Lambda$ are maximized and because gas on very nearly circular  
orbits is a precise probe of the radial force law. For a stable 
circular orbit with velocity $v_c$ at a distance $r$ from the 
center of a galaxy with mass $M$ we obtain using eq. (\ref{acc}) 
\be v_c^2 = {{GM}\over r} -{{\Lambda c^2 r^2}\over 3}  
\label{rotvel} \ee We now define the rescaled dimensionless 
quantities  ${\bar v}_{100}$, ${\bar r}_{10}$ and ${\bar M}_{10}$ 
as follows:  
\begin{equation}
\eqalign{ 
 v_c &= 100 \; {\bar v}_{100} \; km/sec \cr
 r &= 10 \;{\bar r}_{10}\; kpc \cr
 M &= 10^{10} {\bar M}_{10} \; M_\odot
\label{rescaled2} } 
\end{equation}
Eq. (\ref{rotvel}) may now be written in a rescaled form as \be 
{\bar v}_{100}^2 ={1\over 2} {{\bar M}_{10} \over {\bar r}_{10}}- 
3 \times 10^{-5} {\bar \Lambda}_{52}\; {\bar r}_{10}^2  
\label{rescv}\ee In order to calculate the effects of the 
cosmological constant on the dynamically obtained masses of 
galaxies (including their halos) it is convenient to calculate the 
ratio 
%\be p\equiv {{{\bar M}_{10} ({\bar \Lambda}_{52} = 1) -  
%{\bar M}_{10} ({\bar \Lambda}_{52} = 0)}\over {{\bar M}_{10} 
%({\bar \Lambda}_{52} = 0)}}= 
% {{3  \times 10^{-5} {\bar r}_{10}^2}
% \over {{\bar v}_{100}^2}}  \label{massrat}  \ee
  
\be p \equiv {{M ({\bar \Lambda}_{52} = 1) -  M({\bar 
\Lambda}_{52} = 0)}\over {M({\bar \Lambda}_{52} = 0)}}= {{3  
\times 10^{-5} {\bar r}_{10}^2} \over {{\bar v}_{100}^2}}  
\label{massrat}  \ee  

 In Table 1 we show a calculation of the mass ratio $p$ for 22 
galaxies of different sizes and rotation 
velocities\cite{Sanders96}. The corresponding plot of $p(r)$ is 
shown in Fig. 1. 
\begin{figure}[bp1]
 \centerline{\epsfxsize = 0.86\hsize \epsfbox{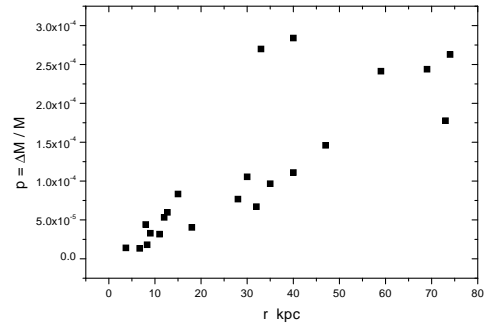}}
\caption[FIG.1]{Relative increase $p$ of dynamically calculated 
mass of galaxies due to repusive effects vacuum energy vs galaxy 
radious $r$ measured in kpc.} \label{fig1} 
\end{figure}
It is clear that even for large galaxies where the role of the 
repulsive force induced by vacuum energy is maximized, the 
increase of the mass needed to compensate vacuum energy 
antigravity is negligible. In order for these effects to be 
significant the cosmological constant would have to be larger that 
the value required for flatness by a factor of at least $10^3$. 
 \begin{table}
 \caption{Relative increase $p$ of dynamically calculated mass of 
 galaxies due to repulsive effects of vacuum energy.}
 \begin{tabular}{lccc} % In second brace, l = left, r = right,
% c = centered and d = decimal justification.
 {\bf Galaxy} & $ r_{HI}$  kpc&$v_{rot}$ km/sec & $p$ 
 %\times 10^{-5}{\bar\Lambda}_{52}$ 
 \\  
 % Separate items with &. End line with \\
 \tableline % Creates a horizontal line.
 %UGC2885&73&300&Four\\ % Place \tablenote{}
% after item to be footnoted.
 UGC2885&73&300&$17.8\times 10^{-5}$\\
 NGC5533&74&250&$26.3\times 10^{-5}$\\
 NGC6674&69&242&$24.3\times 10^{-5}$\\
 NGC5907&32&214&$6.7\times 10^{-5}$\\
 NGC2998&47&213&$14.6\times 10^{-5}$\\
 NGC801&59&208&$24.1\times 10^{-5}$\\
 NGC5371&40&208&$11.1\times 10^{-5}$\\
 NGC5033&35&195&$9.7\times 10^{-5}$\\
 NGC3521&28&175&$7.7\times 10^{-5}$\\
 NGC2683&18&155&$4.0\times 10^{-5}$\\
 NGC6946&30&160&$10.5\times 10^{-5}$\\
 UGC128&40&130&$28.4\times 10^{-5}$\\
 NGC1003&33&110&$27.0\times 10^{-5}$\\
 NGC247&11&107&$3.1\times 10^{-5}$\\
 M33&8.3&107&$1.8\times 10^{-5}$\\
 NGC7793&6.7&100&$1.3\times 10^{-5}$\\
 NGC300&12.7&90&$6.0\times 10^{-5}$\\
 NGC5585&12&90&$5.3\times 10^{-5}$\\
 NGC2915&15&90&$8.3\times 10^{-5}$\\
 NGC55&9&86&$3.2\times 10^{-5}$\\
 IC2574&8&66&$4.4\times 10^{-5}$\\
 DDO168&3.7&54&$1.4\times 10^{-5}$\\ 
 \end{tabular}
 \end{table}
Such value would be inconsistent with several cosmological 
observations even though it is consistent\cite{Whitehouse:1999rs} 
with theoretical expectations based on the Extended Large Number 
Hypothesis. Therefore, even though the effects of vacuum energy on 
galactic dynamics are much more important compared to the 
corresponding effects on solar system dynamics it is clear that we 
must consider systems on even larger scales where the mean density 
is smaller in order to obtain any nontrivial effects on the 
dynamics. 

The Local Group of galaxies is a particularly useful system for 
studying mass dynamics on large scales because it is close enough 
to be measured and modeled in detail yet it is large enough (and 
poor enough) to probe the effects of vacuum energy on the 
dynamics. The dominant members of the group are the Milky Way and 
the Andromeda Nebula M31. Their separation is \be r_0 \equiv r 
(t=t_0) \simeq 800 kpc  \label{galsep}\ee and the rate of change 
of their separation is  \be {{dr}\over dt}(t=t_0) \simeq -123 km 
s^{-1}  \label{galvel}\ee A widely used assumption is that the 
motion of approach of M31 and Milky Way is due to the mutual 
gravitational attraction of the masses of the two galaxies. 
Adopting the simplest model of the Local Group as an isolated two 
body system, the Milky Way and M31 have negligible relative 
angular momentum and their initial rate of change of separation is 
zero in comoving coordinates. The equation of motion for the 
separation $r(t)$ of the centers of the two galaxies in the 
presence of a nonzero cosmological constant is: 
\be
{{d^2 r}\over {dt^2}}= -{{G M}\over r^2} + {{\Lambda c^2}\over 3} 
r  \label{galacc}\ee  where M is the sum of the masses of the two 
galaxies. A similar equation (with $\Lambda = 0$) was used in Ref 
\cite{p93} to obtain an approximation of the mass to light ratio 
of the galaxies of the Local Group. Numerical studies\cite{pmhj89} 
have shown that this approximation is reasonable and leads to a 
relatively small overestimation (about 25\%) of the galactic 
masses. This correction is due to the effects of the other dwarf 
members of the Local Group that are neglected in the isolated two 
body approximation. Here we are not interested in the precise 
evaluation of the masses of the galaxies but on the effects of the 
cosmological constant on the evaluation of these masses. Therefore 
we will use the `isolated two body approximation' of the Local 
Group (eq. (\ref{galacc})) and focus on the dependence of the 
calculated value of the mass $M$ as a function of $\Lambda$ in the 
range of cosmologicaly interesting values of $\Lambda$.  Our goal 
is to find the total mass $M$ of the Local Group galaxies, using 
eq. (\ref{galacc}) supplied with the following conditions: \bea  
r(t=t_0) &=& 800\; kpc \label{condition1}\\ 
 {{dr}\over {dt}}(t=t_0) &=& -123 \; km/sec \label{condition2} \\
{{dr}\over {dt}}(t=0) &=& 0 \label{condition3} \eea  where $t_0 =  
15Gyr$. Upon integrating and rescaling eq. (\ref{galacc}) we 
obtain \be ({{d{\bar r}_{100}}\over {d{\bar t}_{15}}})^2 = {\bar 
M} ({1\over {{\bar r}_{100}}} - {1\over 8}) + {\bar \Lambda} 
({\bar r}_{100}^2 -64) + 420 \equiv f({\bar 
r}_{100})\label{resclg} \ee where we have used condition 
(\ref{condition2}) and the rescaled quantities defined by \bea  r 
&=& 100 \; {\bar r}_{100} \;  kpc \\ t &=& 1.5 \times 10^{10}\; 
{\bar t}_{15} \; yrs \\ M &=& 4\times 10^{8}\; {\bar M}\;  
M_\odot\\ \Lambda &=& 1.3 \; {\bar \Lambda}_{52} 
\label{rescalings} \eea Using now conditions (\ref{condition1}) 
and (\ref{condition3}) we obtain the equation that can be solved 
to evaluate the galactic masses for various $\Lambda$ \be 1 ={\bar 
t}_{15} (t=t_0) = - \int_{{\bar r}_{100}(t=0)}^{{\bar 
r}_{100}(t=t_0)} {{dr}\over \sqrt{f({\bar r}_{100})}} 
\label{intlg} \ee The lower limit of the integral (\ref{intlg}) is 
obtained by solving condition (\ref{condition3}) for $r$ (using 
eq. (\ref{resclg}) while the upper limit is given by eq. 
(\ref{condition1}) in its rescaled form. This equation can be 
solved numerically for $M$ to calculate the galactic total mass 
$M$ for various values of the cosmological constant $\Lambda$.  
The resulting dependence of $M$ on $\Lambda$ is shown in Fig. 2 
(continuous line). 
\begin{figure}[bp2]
 \centerline{\epsfxsize = 0.86\hsize \epsfbox{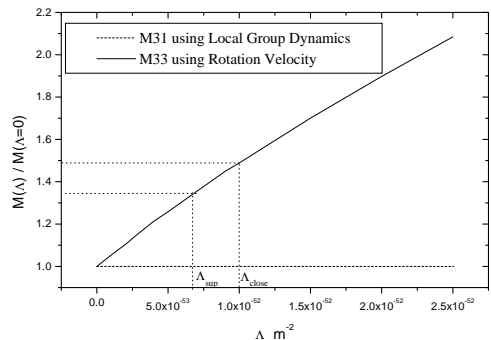}}
\caption[FIG.2]{The relative galactic total mass 
${{M(\Lambda)}\over {M(\Lambda=0)}}$ calculated for various values 
of the cosmological constant $\Lambda$ using Local Group 
(continous line) and galactic scale (M33, dashed line) velocity 
data. The dotted lines correspond to the $\Lambda$ value implied 
by the SnIa data and to $\Lambda_{cl}$ for which the vacuum energy 
alone closes the universe.} \label{fig2} 
\end{figure}

Clearly, for a value of $\Lambda$ consistent with the recent SnIa 
observations ($\Lambda\simeq 0.7\times 10^{-52}m^{-52}$) the 
calculated galactic masses using Local Group dynamics are 35\% 
larger than the corresponding masses calculated with $\Lambda = 
0$. On Fig. 2 we also plot (dashed line) the dependence of the 
${{M(\Lambda)}\over {M(\Lambda = 0)}} $ on $\Lambda$ calculated 
using galactic dynamics (rotation velocity) of the galaxy M33. 
Clearly a galactic system in contrast to the Local Group is too 
small and dense to be a sensitive detector of the cosmological 
constant.  

\section{Conclusion} 
We conclude that the Local Group of galaxies is a system that is 
large enough and with low enough matter density to be a sensitive 
probe of the gravitational effects of a cosmological constant with 
value consistent with cosmological expectations and recent SnIa 
obsrevations. Smaller and denser systems do not have this 
property. On the other hand the gravitational effects of $\Lambda$ 
should be even more pronounced on larger cosmological systems. 

The specific cluster dynamic effect which is discussed here offers 
an independent determination of the cosmological constant 
contribution to a galactic `dark matter' halo. This is consistent 
with the fact that galactic masses of galaxies like our Milky Way 
can be deduced from cluster dynamics such as our Local Group as 
well as by using non-dynamical methods (e.g. gravitational 
lensing).

It is legitimate to wonder whether the distinctive action of the 
cosmological constant over small versus larger scales persists if 
a different form of vacuum energy such as 
`quintessence'\cite{quint} is dominant. This type of effective 
`scalar matter' possesses an equation of state $p=w\rho$ 
($-1<w<0$) and is much softer than the one associated with the 
cosmological constant ($w=-1$). As such it interpolates between 
the latter a normal dark matter component ($w\geq 0$). Moreover we 
should expect that it causes a much smaller repulsion effect 
smoothly passing over to a purely dark matter for increasing $w$. 
As a consequence our analysis for such a `reduced' type of 
antigravity would most probably imply a successively smaller 
discrepancy for the relative total galactic mass contribution of 
quintessence as derived from the Local Group versus the galactic 
velocity data for increasing $w$ ($-1 < w < 0$). A recent analysis 
of the dynamical effects of quintessence on flat galactic rotation 
curves\cite{matos} in fact corroborates to this point of view.   

The gravitational effects discussed here can only be used as an 
independent detection method of the cosmological constant, if the 
galactic masses of systems like the Local Group are measured 
independently using non-dynamical methods (eg gravitational 
lensing).  In that case $\Lambda$ could be obtained using plots 
like the one shown in Fig. 2. This type of investigation is 
currently in progress. 

\section{Acknowledgements}

We would like to thank M. Plionis for useful conversation.

%\bibliographystyle{prsty}

%\bibliography{bibliog}

\begin{thebibliography}{10}

\bibitem{scpc99}
S. Perlmutter et al. [Supernova Cosmology Project Collaboration], 
Ap. J. {\bf 517}, 565 (1999); astro-ph/9812133. 
\bibitem{hz98}
B. P. Schmidt et al. [Hi-Z Supernova Team Collaboration], 
Astrophys. Journ. 507, 46 (1998); astro-ph/9805200; 
%\cite{Riess:1998cb}
\bibitem{Riess:1998cb}A.~G.~Riess {\it et al.},
%``Observational Evidence from Supernovae for an Accelerating Universe and a Cosmological Constant,''
Astron.\ J.\  {\bf 116}, 1009 (1998)[astro-ph/9805201]. 
%\cite{Weinberg:1996xe}
\bibitem{Weinberg:1996xe}S.~Weinberg,
``Theories of the cosmological constant,'' astro-ph/9610044. 
%%CITATION = ASTRO-PH 9610044;%%
%\href{http://www.slac.stanford.edu/spires/find/hep/www?eprint=ASTRO-PH/9610044}{SPIRES}
%\cite{Krauss:1995yb}
\bibitem{Krauss:1995yb}
L.~M.~Krauss and M.~S.~Turner, 
%``The cosmological constant is back,''
Gen.\ Rel.\ Grav.\  {\bf 27}, 1137 (1995) [astro-ph/9504003]. 
%%CITATION = ASTRO-PH 9504003;%%
%\href{http://www.slac.stanford.edu/spires/find/hep/www?eprint=ASTRO-PH/9504003}{SPIRES}
%\cite{Carroll:1992mt}
\bibitem{Carroll:1992mt}
S.~M.~Carroll, W.~H.~Press and E.~L.~Turner, 
%``The Cosmological constant,''
Ann.\ Rev.\ Astron.\ Astrophys.\  {\bf 30}, 499 (1992). 
%%CITATION = ARAAA,30,499;%%
%\href{http://www.slac.stanford.edu/spires/find/hep/www?j=ARAAA%2c30%2c499}{SPIRES}
%\cite{Ostriker:1995su}
\bibitem{Ostriker:1995su}
J.~P.~Ostriker and P.~J.~Steinhardt, 
%``The Observational case for a low density universe with a non-zero cosmological constant,''
Nature {\bf 377}, 600 (1995). 
%\cite{Quast:1999fh}
\bibitem{Quast:1999fh}R.~Quast and P.~Helbig,
%``Gravitational lensing statistics with extragalactic surveys. I. A lower limit on the cosmological constant,''
Astron.\ Astrophys.\  {\bf 344}, 721 (1999)[astro-ph/9904174]. 
%\cite{Zehavi:1999fm}
\bibitem{Zehavi:1999fm}I.~Zehavi and A.~Dekel,
%``Constraints on the Cosmological Constant from Flows and Supernovae,''
astro-ph/9904221. 
\bibitem{Sanders96}
R.H. Sanders, Ap. J. {\bf 473}, 117  (1996). 
%\cite{Whitehouse:1999rs}
\bibitem{Whitehouse:1999rs}
S.~B.~Whitehouse and G.~V.~Kraniotis, 
%``A possible explanation of Galactic Velocity Rotation Curves in terms of a Cosmological Constant,''
astro-ph/9911485. 
\bibitem{ct98} J. Cardona and J. Tejeiro, Ap. J. {\bf 493}, 52 (1998).
%\cite{Wright:1998bc}
\bibitem{Wright:1998bc}
E.~L.~Wright, 
%``Interplanetary Measures Can Not Bound the Cosmological Constant,''
astro-ph/9805292. 
%%CITATION = ASTRO-PH 9805292;%%
%\href{http://www.slac.stanford.edu/spires/find/hep/www?eprint=ASTRO-PH/9805292}{SPIRES}
%\cite{Neupane:1999hr}
\bibitem{Neupane:1999hr}I.~P.~Neupane,
%``Planetary perturbation with cosmological constant,''
gr-qc/9902039. 
%\cite{Roberts:1987ch}
\bibitem{Roberts:1987ch}
M.~D.~Roberts, 
%``The Orbit Of Pluto And The Cosmological Constant,''
Mon.\ Not.\ Roy.\ Astron.\ Soc.\  {\bf 228}, 401 (1987). 
%%CITATION = MNRAA,228,401;%%
%\href{http://www.slac.stanford.edu/spires/find/hep/www?j=MNRAA%2c228%2c401}{SPIRES}
%\cite{White:1998fn}
\bibitem{White:1998fn}M.~White,
%``Complementary Measures of the Mass Density and Cosmological Constant,''
Astrophys.\ J.\  {\bf 506}, 495 (1998)[astro-ph/9802295]. 
%\cite{Zlatev:1999tr}
\bibitem{Zlatev:1999tr}
I.~Zlatev, L.~Wang and P.~J.~Steinhardt, 
%``Quintessence, Cosmic Coincidence, and the Cosmological Constant,''
Phys.\ Rev.\ Lett.\  {\bf 82}, 896 (1999)[astro-ph/9807002]. 
%\cite{Garriga:1999bf}
\bibitem{Garriga:1999bf}J.~Garriga and A.~Vilenkin,
%``On likely values of the cosmological constant,''astro-ph/9908115.
%\cite{Garriga:2000hu}\bibitem{Garriga:2000hu}
J.~Garriga, M.~Livio and A.~Vilenkin, 
%``The cosmological constant and the time of its dominance,''
Phys.\ Rev.\  {\bf D61}, 023503 (2000)[astro-ph/9906210].  
%\cite{Gibbons:1977mu}
\bibitem{Gibbons:1977mu}
G.~W.~Gibbons and S.~W.~Hawking, 
%``Cosmological Event Horizons, Thermodynamics, And Particle Creation,''
Phys.\ Rev.\  {\bf D15}, 2738 (1977). 
%%CITATION = PHRVA,D15,2738;%%
%\href{http://www.slac.stanford.edu/spires/find/hep/www?j=PHRVA%2cD15%2c2738}{SPIRES}
\bibitem{p93}
P.J.E. Peebles, {\it Principles of Physical Cosmology}, Princeton 
Series in Physics (1993). 
\bibitem{pmhj89}
P.J.E. Peebles, A.L. Melott, M.R. Holmes and L.R. Jiang, Ap. J.  
{\bf 345}, 108 (1989). 
\bibitem{quint}
R.~R.~Caldwell, R.~Dave and P.~J.~Steinhardt, 
%``Cosmological Imprint of an Energy Component with General Equation-of-State,''
Phys.\ Rev.\ Lett.\  {\bf 80}, 1582 (1998) [astro-ph/9708069]; 
I.~Zlatev, L.~Wang and P.~J.~Steinhardt, 
%``Quintessence, Cosmic Coincidence, and the Cosmological Constant,''
Phys.\ Rev.\ Lett.\  {\bf 82}, 896 (1999) [astro-ph/9807002].  
\bibitem{matos}
T.~Matos and F.~S.~Guzman, 
%``Quintessence at galactic level?,''
astro-ph/0002126; F.~S.~Guzman, T.~Matos, D.~Nunez and E.~Ramirez, 
%``Quintessence-like dark matter in spiral galaxies,''
astro-ph/0003105.  
\end{thebibliography}

\end{document}